\documentclass[12pt]{article}
\usepackage{epsfig}
\usepackage{cite}

\def\beq{\begin{equation}}
\def\eeq#1{\label{#1}\end{equation}}
\def\eeqn{\end{equation}}

\def\beqa{\begin{eqnarray}}
\def\eeqa#1{\label{#1}\end{eqnarray}}
\def\eeqan{\end{eqnarray}}

\let\bar=\overbar

\def\Dslash{\not{\hbox{\kern-4pt $D$}}}
\def\dslash{\not{\hbox{\kern-2pt $\del$}}}

\def\msb{{\bar{\ssstyle M \kern -1pt S}}}

\def\Title#1{\begin{center} {\Large {\bf #1} } \end{center}}

\begin{document}

\Title{Infrared Divergences from Soft and Collinear Gauge Bosons}

\bigskip\bigskip


\begin{raggedright}

{\it Paul Jameson\index{Jameson, P.}\\
School of Mathematics and Statistics\\
University of Plymouth\\
Plymouth, PL4 8AA, U.K.}
\bigskip\bigskip
\end{raggedright}

\interfootnotelinepenalty=10000

\section{Introduction}

To compare with experiment it is vital that theoretical predictions are finite. Beyond leading order in perturbation theory many processes contain infrared divergences. It is often argued that these singularities are eliminated at the level of the inclusive cross-section \cite{Bloch:1937pw,Lee:1964is}. During this talk I will show that the infrared divergences are still poorly understood. I also expose some questionable assumptions which must be made to render the theoretical cross-section finite.

There are two types of infrared singularities -- soft and collinear. Soft divergences are associated with low energy photons. Collinear divergences occur in high-energy or theories with massless charges. It is common practice to use the Bloch-Nordsieck trick \cite{Bloch:1937pw} for the soft divergences. By adding the bremsstrahlung process to the cross-section with a virtual loop the overall cross-section is soft finite. Soft divergences can be regulated by dimensional regularisation where the number of dimensions, $D=4+2 \varepsilon_\mathrm{IR}$. Focusing on the infrared pole, the Bloch-Nordsieck trick is summarised diagrammatically at next-to-leading order by:
\begin{equation}
\label{BN}
 2\,\left(\parbox{20mm}{\includegraphics[scale=0.4]{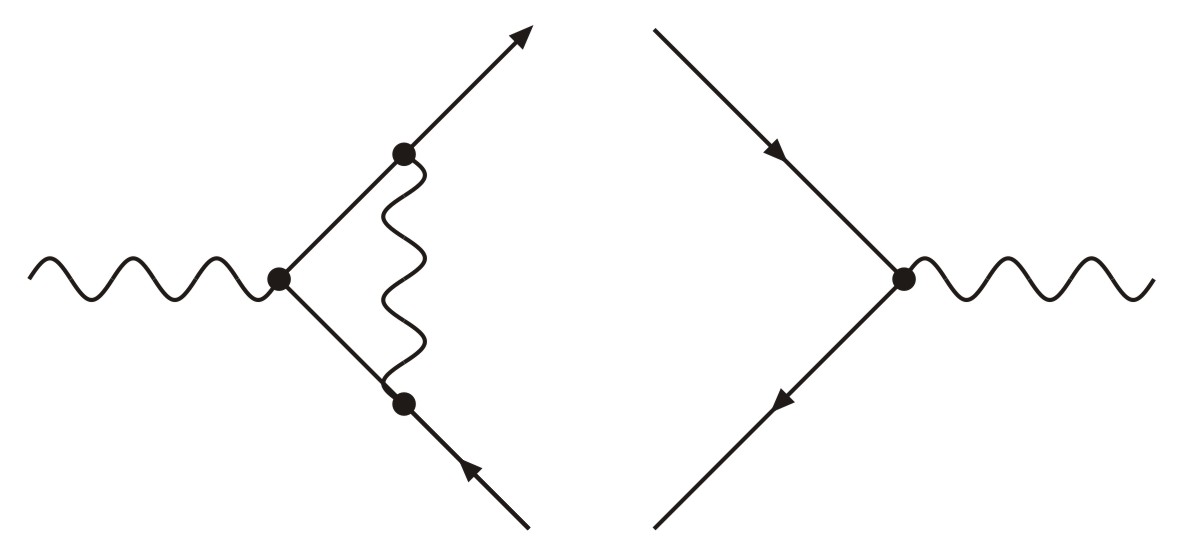}}
\qquad\qquad\quad\right)+
\left|\parbox{20mm}{\includegraphics[scale=0.4]{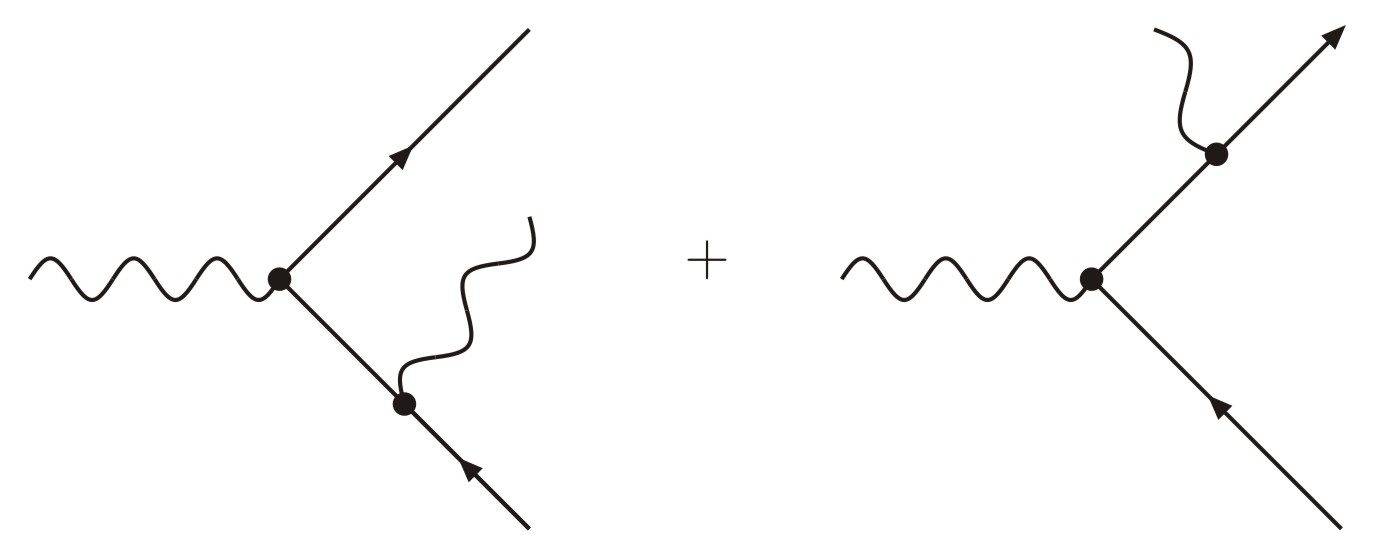}
}\qquad\qquad\qquad \, \,\right|^2  =
-\frac{1}{\varepsilon_\mathrm{IR}}+\frac{1}{\varepsilon_\mathrm{IR}}\,,
\end{equation}
up to an overall multiplicative constant.
This sum corresponds to the idea that there may be low energy unobserved photons accompanying charges in the final state. Such processes are indistinguishable by experiment from an isolated charge. They are referred to as degenerate processes. The Bloch-Nordsieck trick, however, does not work for collinear divergences.

The standard approach to deal with collinear divergences was developed by Lee and Nauenberg \cite{Lee:1964is}. Their quantum mechanical theorem states that the sum over all degenerate processes is infrared finite. In field theory one should thus include \emph{both} initial and final state degeneracies. In this talk I will analyse the Lee-Nauenberg theorem applied to high-energy QED. Collinear divergences will be regulated by the effectively small mass, $m$, of the electron and will appear as terms proportional to $\ln(m)$.

\section{New Class of Collinear Divergences}

It is assumed in \cite{Lee:1964is} that soft divergences have been dealt with by the Bloch-Nordsieck trick and thus only collinear singularities produced by photons with energies greater than the experimental resolution $\Delta$ are considered.
Recent work has shown that there is a class of divergences which were omitted in Lee and Nauenberg's paper \cite{Lavelle:2005bt}. These come from processes where low energy (soft) photons travelling parallel (collinearly) to the charges are emitted and/or absorbed. These processes contain $\Delta \ln (m)$ collinear singularities in their cross-section which, from now on, will be referred to as $\Delta$--divergences. By inserting these divergences into the analysis conducted in \cite{Lee:1964is} it may be seen that the cross-section cannot be simultaneously both soft and collinear finite. The reason for this failure may be traced back to the different ways soft and collinear singularities are dealt with.

The Lee-Nauenberg theorem states that one should consider initial state degeneracies. However, at first sight, including the absorption of low energy photons has the effect of reintroducing the soft divergences so that (\ref{BN}) becomes:
\begin{eqnarray}
\label{BN + abs}2\,\left(\parbox{20mm}{\includegraphics[scale=0.4]{vir_x_sec.jpg}}
\qquad\qquad\qquad\right)+
\left|\parbox{20mm}{\includegraphics[scale=0.4]{emission_no_labels.jpg}
}\qquad\qquad\qquad\,\,\,\right|^2 +
\\ \nonumber \left|\parbox{20mm}{\includegraphics[scale=0.4]{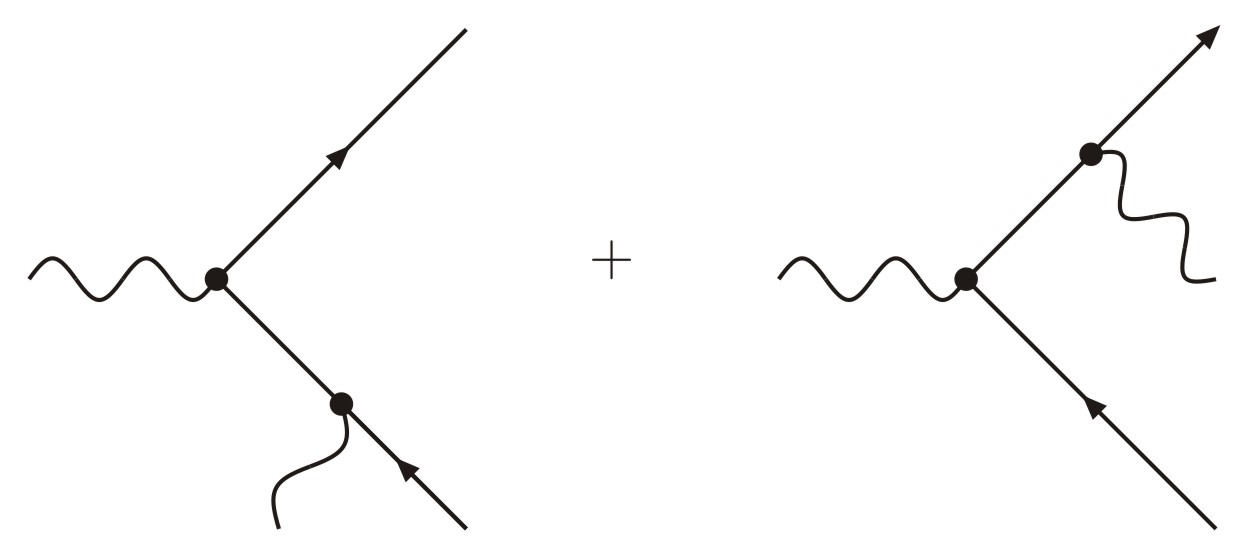}
}\qquad\qquad\qquad\right|^2 =
   -\frac{1}{\varepsilon_\mathrm{IR}}+\frac{1}{\varepsilon_\mathrm{IR}}+ \frac{1}{\varepsilon_\mathrm{IR}}\,.
\end{eqnarray}
However, all degeneracies must be considered and Lavelle and McMullan have shown that the residual soft divergences in (\ref{BN + abs}) may be cancelled by including the following cross-section contributions:
\begin{equation}
\label{ML DM}
2\,\left(\parbox{20mm}{\includegraphics[scale=0.4]{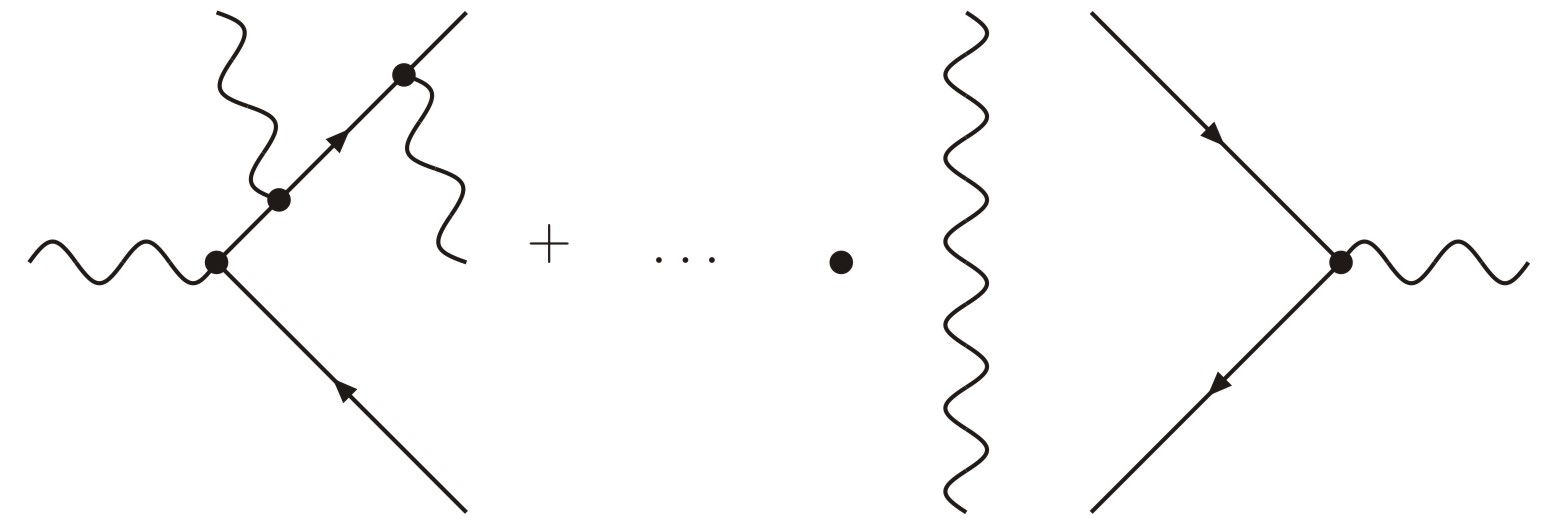}}
\qquad\qquad\qquad\qquad\right)+\left|\parbox{20mm}{\includegraphics[scale=0.4]{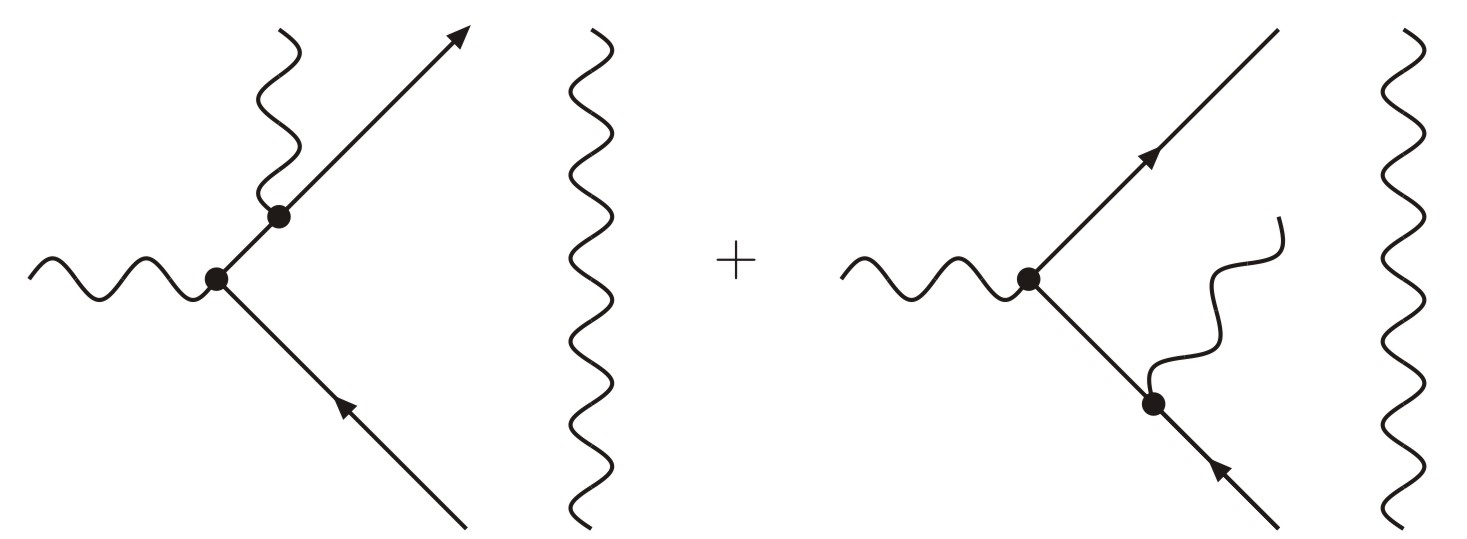}}
\qquad\qquad\qquad\qquad\right|^2\,= -\frac{2}{\varepsilon_{\mathrm{IR}}} +\frac{1}{\varepsilon_{\mathrm{IR}}}\,.
\end{equation}
The striking thing here is the need for the disconnected processes in the application of the Lee-Nauenberg theorem. This was only briefly noted in the original paper \cite{Lee:1964is}.
Investigations, by a variety of authors \cite{Lavelle:2005bt,DeCalan:1972ya,Ito:1980jt,Muta:1981pe,Axelrod:1985yi,Akhoury:1997pb}, have shown the need to consider such disconnected processes as displayed above.
The amplitudes which interfere with the disconnected ones are the emission and absorption of a photon which can take place with either electron leg.
By including all the above processes the $\ln(m)$ and soft singularities are eliminated. However, in order to fully determine whether the cross-section is finite the $\Delta$--divergences must also be studied. The aim of this work is to discover if the Coulomb scattering cross-section can be made simultaneously both soft and collinear infrared finite.

Lee and Nauenberg's theorem states that one should include all degenerate processes at each order of perturbation theory. However, at the same order of perturbation theory an infinite number of disconnected photons may be considered! Processes contributing to the next-to-leading order Coulomb scattering cross-section may be clustered into groups in which the soft singularities are eliminated:
\begin{equation}\label{bn10}
  \frac{1}{\varepsilon_\mathrm{IR}}\left[ \underbrace{-1+1}_{\mathrm{Bloch-Nordsieck}}+\,\,\underbrace{(1-2+1)}_{\mathrm{Lee-Nauenberg}}\,\,+\,\,\underbrace{(1-2+1)}_{\mathrm{Lee-Nauenberg}}\,\,+
  \,\,\underbrace{(1-2+1)}_{\mathrm{Lee-Nauenberg}}\,\,+\,\,\dots\right]\,.
\end{equation}
For each additional disconnected photon there are three further processes to consider. By calculating the soft divergences one obtains another factor of $(1-2+1)$. Higher numbers of disconnected photons are \emph{not} suppressed in this series so it will not converge. In practice this series is arbitrarily truncated at one of the finite stages to obtain a finite cross-section. Further discussion of this mathematically ill-defined procedure may be found in \cite{Lavelle:2005bt}.

\section{Cancellation of $\Delta$--Divergences}

\begin{figure}[htb]
\begin{center}
\includegraphics[scale=0.6]{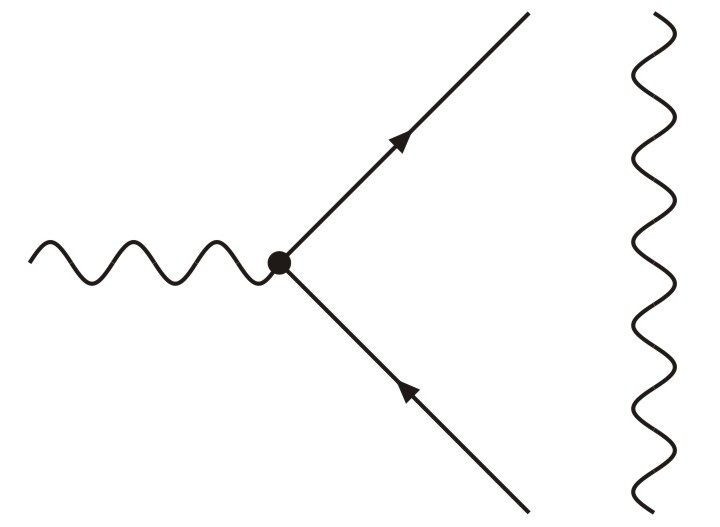}
\caption{A process involving a disconnected photon.}
\label{fig:disc}
\end{center}
\end{figure}

The disconnected photon, displayed in Figure \ref{fig:disc}, cannot be collinear with the electron both before and after it has been scattered. Therefore, in order for it to contribute to Coulomb scattering (and not be identified as a different process) the photon's energy must be lower than the experiment's detector resolution. Disconnected processes provide contributions to the cross-section which can be soft \emph{and} collinear. Therefore they produce $\Delta$--divergences. The question arises: do these $\Delta$ divergences cancel for the same combination of diagrams as the soft singularities (\ref{bn10})?

Photons may be detected by an experiment in two ways:
\begin{enumerate}
\item Directly, through a detector with resolution $\Delta$;
\item Indirectly, through a measurement of the electron's `missing' energy.
\end{enumerate}
A shift in the electron's energy puts an upper limit on the total energy contained by any number of soft photons. An electron energy detector is a powerful tool for determining whether a process is different from tree-level Coulomb scattering if multiple photons are contributing to the process. If the process contributes to Coulomb scattering then the photon's energy (or energies when including disconnected photons) is such that it cannot be detected by either of the above methods.

I have studied the next-to-leading order Coulomb scattering cross-section and evaluated all the $\Delta$--divergences. The technical details will not be presented here (see \cite{me}) and instead I will state the results. Removal of $\Delta$--divergences from the cross-section may take place using exactly the same prescription as soft divergences (\ref{bn10}). However, for this cancellation to take place the following assumptions are required.
\begin{enumerate}
\item All processes must have an inverted energy weighting\footnote{I have followed the lead set by Bethe and Heitler. They state that for the emission
cross-section an \emph{energy weighting} factor which is equal to the electron energy in the out-state divided by the electron energy in the in-state should be included. For references see section 5-2-4 in \cite{Itzykson:1980rh},  p. 499 of \cite{Brown:1992db}, p. 244 of \cite{Heitler}, p. 309 of \cite{Bohm:2001yx} and the original paper \cite{Bethe-Heitler}. Intuitively the probability for emission and absorption should be equivalent. This requirement leads to the introduction of an \emph{inverted} energy weighting for the absorption process. In the literature the energy weighting has only been carefully considered for the emission process. Collinear $\Delta$--divergences are highly sensitive to the energy weighting scheme used so it will be important while studying their cancellation.}. The exception to this rule is the emission process for which the usual Bethe-Heitler weighting is applied.
\item Experimentalists \emph{must} observe photons \emph{directly} through a photon detector i.e. they are \emph{not} allowed to indirectly observe a photon through `missing' electron energy.
\end{enumerate}
The energy weighting applied is not supported by a physical motivation (other than the infrared finiteness which I am trying to prove). However, without this choice of weighting the Lee-Nauenberg theorem fails because the $\Delta$--divergences are sensitive to the energy weighting. My calculations prove that in order for the $\Delta$--divergences to cancel, the maximum photon energy must be the same for each of the processes being combined. If a photon is detected indirectly then the amplitudes which contain different numbers of soft photons will have different maximum photon energies. The processes required to cancel the soft singularities, displayed in equations (\ref{BN + abs}) and (\ref{ML DM}), contain a different number of soft photons so the second assumption is necessary.

Weinberg has made his reservations clear on the application of the Lee-Nauenberg theorem. I quote from page 552 of \cite{Weinberg:1995mt} \emph{``$\dots$ to the best of my knowledge no one has given a complete demonstration that the sums of transition rates that are free of infrared divergences are the only ones that are experimentally measurable.''} Through this calculation I have shown that Weinberg's reservations are well-founded since infrared finiteness places a condition on the way experiments are conducted.

In summary what I have seen is that not only does the Lee-Nauenberg approach to the infrared lead to an ill defined series of diagrams but there is no natural way to achieve infrared finiteness. Further research is urgently required.

\bigskip
I am grateful to Martin Lavelle and David McMullan for their advice and assistance during the conduct of this research.


\begin{thebibliography}{99}


\bibitem{Bloch:1937pw}
F.~Bloch and A.~Nordsieck,
\newblock Phys. Rev. {\bf 52}, 54 (1937).

\bibitem{Lee:1964is}
T.~D. Lee and M.~Nauenberg,
\newblock Phys. Rev. {\bf 133}, B1549 (1964).

\bibitem{Lavelle:2005bt}
M.~Lavelle and D.~McMullan,
\newblock JHEP {\bf 03}, 026 (2006), hep-ph/0511314.

\bibitem{DeCalan:1972ya}
C.~De~Calan and G.~Valent,
\newblock Nucl. Phys. {\bf B42}, 268 (1972).

\bibitem{Ito:1980jt}
I.~Ito,
\newblock Prog. Theor. Phys. {\bf 65}, 1466 (1981).

\bibitem{Muta:1981pe}
T.~Muta and C.~A. Nelson,
\newblock Phys. Rev. {\bf D25}, 2222 (1982).

\bibitem{Axelrod:1985yi}
A.~Axelrod and C.~A. Nelson,
\newblock Phys. Rev. {\bf D32}, 2385 (1985).

\bibitem{Akhoury:1997pb}
R.~Akhoury, M.~G. Sotiropoulos, and V.~I. Zakharov,
\newblock Phys. Rev. {\bf D56}, 377 (1997), hep-ph/9702270.

\bibitem{Itzykson:1980rh}
C.~Itzykson and J.~B. Zuber,
\newblock {\em Quantum Field Theory} ,
\newblock McGraw-Hill, New York (1980).

\bibitem{Brown:1992db}
L.~S. Brown,
\newblock {\em Quantum field theory} ,
\newblock Cambridge, UK: Univ. Pr. (1992).

\bibitem{Heitler}
W.~Heitler,
\newblock {\em The Quantum Theory of Radiation} ,
\newblock Oxford University Press, (1950).


\bibitem{Bohm:2001yx}
M.~Bohm, A.~Denner, and H.~Joos,
\newblock Stuttgart, Germany: Teubner (2001).


\bibitem{Bethe-Heitler}
H.~Bethe and W.~Heitler,
\newblock Proc. Roy. Soc. A {\bf 146}, 83 (1934).

\bibitem{me}
P.~Jameson, M.~Lavelle and D.~McMullan,
\newblock in preparation.

\bibitem{Weinberg:1995mt}
S.~Weinberg,
\newblock {\em The Quantum Theory of Fields. Vol. 1: Foundations} ,
\newblock CUP, (1995) 609 p.

\end{thebibliography}
\end{document}